\documentclass[aps,twocolumn,prb,eqsecnum,showpacs]{revtex4}
\usepackage{graphicx}
\begin{document}
\draft
\preprint{16 September 2008}
\title{Optical characterization of ground states of polyacene}
\author{Shoji Yamamoto}
\address{Department of Physics, Hokkaido University,
         Sapporo 060-0810, Japan}
\date{16 September 2008}
\begin{abstract}
We investigate the ground-state properties of polyacene in terms of an
extended Peierls-Hubbard Hamiltonian with particular emphasis on its
structural instability of two types: double bonds in a {\it cis} pattern
and those in a {\it trans} pattern.
Calculating the polarized optical conductivity spectra within and beyond
the Hartree-Fock scheme, we reveal a striking contrast between the
{\it cis} and {\it trans} configurations.
The two Peierls-distorted states are almost degenerate in their
energetics but quite distinct in their optics.
\end{abstract}
\maketitle

\section{Introduction}

   There is a lot of enthusiasm for semiconducting organic materials,
which are available at low cost, easy to process, lightweight, and
flexible.
Especially for the last few decades, particular attention has been
paid to polyacene oligomers, consisting of an aromatic linear array, due
to their multifunctional electronic properties.
Tetracene and pentacene, for instance, serve as light-emitting devices
\cite{S599,B035321} and field-effect transistors. \cite{G87,S1022,B4773}
The recent discovery of graphene \cite{N666} and the success of patterning
it into a narrow ribbon \cite{N197,Z201,B1191} have stimulated a renewed
interest \cite{P255207} in polyacene as a basic building unit of them.

\begin{figure}
\centering
\includegraphics[width=82mm]{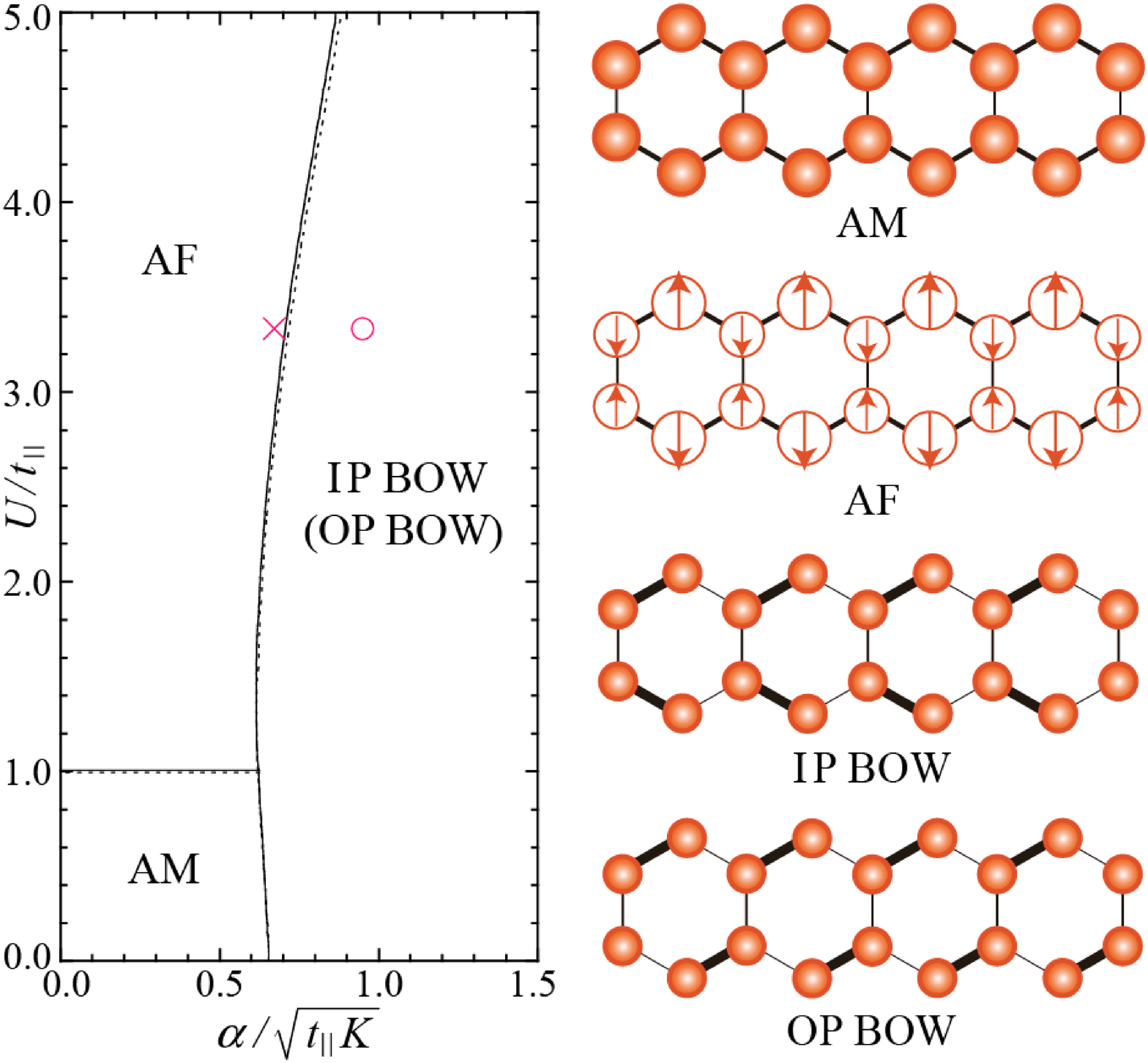}
\vspace*{-2.5mm}
\caption{(Color online)
         HF (dotted lines) and SDCI (solid lines) calculations of a
         ground-state phase diagram on the $\alpha$-$U$ plain at $N=16$
         under the screened-Coulomb parametrization.
         $\times$ and $\circ$ indicate the two sets of model parameters
         considered here.
         We find a metallic state of aromatic (AM) configuration,
         an antiferromagnetic (AF) Mott insulator, and
         {\it cis-} and {\it trans-}distorted Peierls insulators,
         which read as in-phase (IP) and out-of-phase (OP) bond-order-wave
         (BOW) states, respectively.
         All the phases obtained are schematically illustrated, where
         various segments qualitatively represent the variation of local
         bond orders, while arrows in circles depict the alternation of
         local spin densities.
         IP BOW and OP BOW are degenerate with each other within the HF
         scheme, whereas IP BOW is slightly favored over OP BOW in the
         SDCI scheme.}
\label{F:PhD}
\end{figure}

   Prior to the advent of the modern microelectronics, not a few
scientists were already interested in the energy structure of polyacene
from a theoretical point of view.
Early investigations, initiated by Salem and Longuet-Higgins, \cite{S435}
focused on the ground state of polyacene---whether and how the Peierls
distortion occurs.
The discussion was more and more activated with the fabrication of
highly conductive {\it trans-}polyacetylene. \cite{C1098}
There may be a structural instability in polyacene, \cite{S435} as well as
in polyacetylene, \cite{L172} on a qualitatively distinct mechanism,
\cite{K97} however.
Although much effort has so far been devoted to predicting which
structure is energetically preferable in polyacene---{\it cis-}distorted,
{\it trans-}distorted, or uniform (cf. Fig. \ref{F:PhD}),
this long-standing problem is not yet fully settled.
Every structure was actually nominated for the most stable in the
literature. \cite{B4891}
The controversy suggests that all these structures should closely compete
with each other.

   First of all the {\it cis-} and {\it trans-}distorted isomers of
infinite polyacene have exactly the same energy within an electron-phonon
model without any Coulomb interaction \cite{S1269} such as the
Su-Schrieffer-Heeger (SSH) Hamiltonian for polyacetylene. \cite{S1698}
They remain degenerate with each other unless we take account of Coulomb
interactions beyond the Hartree-Fock (HF) scheme.
Secondly the Peierls instability in polyacene is conditional for both
{\it cis} and {\it trans} forms, \cite{S435,R155204} where the
electron-phonon coupling strength seriously affects the scenario.
Thirdly electron correlations make the situation more difficult.
\cite{S8927}
They enhance the Peierls instability on one hand, while they push the
Mott-Hubbard antiferromagnetic insulator on the other hand, against the
uniform aromatic configuration.

   In such circumstances, Ramasesha and coworkers \cite{S8927, R155204}
applied modern numerical tools to polyacene modeled on correlated
Hamiltonians of the Hubbard and Pariser-Parr-Pople (PPP) types, which
include the on-site repulsion alone and the power-law-decaying long-ranged
interaction whole, respectively, but both take no account of spontaneous
phonons.
They claim that the {\it cis} distortion is lower in energy than the
{\it trans} one, both of which may, however, be replaced by a spin density
wave (SDW) on the uniform lattice with increasing Coulomb interactions.
While other authors \cite{H5517,B7416,H134309} predict a fully delocalized
nonalternating nature of polyacene employing standard and developed
density-functional methods, the energy difference among such competing
states must be rather small \cite{B4891} in any case.
Then distorted and undistorted regions may coexist in actual polyacene
compounds due to inevitable local defects and/or possible thermal
excitations.
We are now led to take an interest in characterizing these likely states
rather than ranking them.
Even if there is a slight energy difference between the {\it cis-} and
{\it trans-}distorted structures, how can we tell the lower from the
higher in practice?
We suggest the polarized optical spectroscopy for polyacene in answer to
this question.

   It deserves special mention in the context of our interest that Sony
and Shukla \cite{S155208} have recently calculated the optical absorption
spectra of oligoacenes on a large scale within correlated Hamiltonians of
the PPP type.
Starting from the restricted HF state, which is an undistorted
paramagnetic metal, they investigate in detail configuration interactions
(CIs) between its particle-hole excitations of various levels---single,
double, and more.
In order to describe the effects of electron correlations on the optical
properties of aromatic molecules as precisely as possible, their argument
is restricted to small oligomers up to seven benzene rings and any
possibility of lattice distortion is ruled out.
Other groups of theoreticians \cite{S1269,L77,L477,W341,A2023,A467,Z53}
investigate an infinite acene chain within uncorrelated models of the SSH
type in an attempt to find nonlinear excitations such as solitons and
polarons, where any scenario is constructed on a Peierls-distorted
structure without any concern for its configuration---{\it cis} or
{\it trans}---because of the degenerate energetics.
Thus and thus, there is, to our knowledge, no guiding calculation of the
optical properties of polyacene under both electron-electron and
electron-lattice interactions adequately existent.

\section{Modeling and Formulation}

   We aim at revealing the generic behavior of polyacene rather than
listing individual features of small oligoacenes.
In order to simulate an infinite chain efficiently, we adopt the periodic
boundary condition and restrict Coulomb interactions to a certain range.
Srinivasan {\it et al.} \cite{S8927} investigated the ground-state
properties of polyacene applying a projector Monte Carlo method to the
Hubbard Hamiltonian with periodic boundaries, while Raghu {\it et al.}
\cite{R155204} carried out similar calculations applying a
density-matrix renormalization-group method to the PPP Hamiltonian with
open boundaries.
They declare that the long-range nature of electron-electron interactions
does not qualitatively affect their essential findings such as the
conditional Peierls instability and the {\it cis} distortion favored over
the {\it trans} one.
Sony and Shukla \cite{S155208} further demonstrate that even the H\"uckel
modeling succeeds in reproducing some aspects of the linear optical
spectra of fully correlated oligoacenes modeled on the PPP Hamiltonian.
The low-lying optical absorptions are well describable with both H\"uckel
and PPP models, though the H\"uckel scheme considerably underestimates the
optical gap in general.

   Thus convinced, we depict polyacene in terms of an extended
Peierls-Hubbard Hamiltonian:
\begin{eqnarray}
   &&\!\!\!\!\!\!\!
   {\cal H}=
   -\sum_{l=1}^2\sum_{n=1}^N\sum_{s=\pm}
    \big[
     (t_{\parallel}-\alpha r_{l:2n-1})
     c_{l:2n-1,s}^\dagger c_{l:2n,s}
   \nonumber \\
   &&\!\!\!\!\!\!\!\qquad
    +(t_{\parallel}-\alpha r_{l:2n})
     c_{l:2n,s}^\dagger c_{l:2n+1,s}
    +{\rm H.c.}
    \big]
   \nonumber \\
   &&\!\!\!\!\!\!\!\qquad
   -t_{\perp}\sum_{n=1}^N\sum_{s=\pm}
     \big(c_{1:2n-1,s}^\dagger c_{2:2n-1,s}+{\rm H.c.}\big)
   \nonumber \\
   &&\!\!\!\!\!\!\!\qquad
   +U\sum_{l=1}^2\sum_{n=1}^N
    \Bigl[
     \Bigl(n_{l:2n-1,+}-\frac{1}{2}\Bigr)
     \Bigl(n_{l:2n-1,-}-\frac{1}{2}\Bigr)
   \nonumber \\
   &&\!\!\!\!\!\!\!\qquad
    +\Bigl(n_{l:2n  ,+}-\frac{1}{2}\Bigr)
     \Bigl(n_{l:2n  ,-}-\frac{1}{2}\Bigr)
    \Bigr]
   +V_{\parallel}\sum_{l=1}^2\sum_{n=1}^N\sum_{s,s'=\pm}
   \nonumber \\
   &&\!\!\!\!\!\!\!\qquad\times
    \Bigl(n_{l:2n,s}-\frac{1}{2}\Bigr)
    (n_{l:2n-1,s'}+n_{l:2n+1,s'}-1)
   \nonumber \\
   &&\!\!\!\!\!\!\!\qquad
   +V_{\perp}\sum_{n=1}^N\sum_{s,s'=\pm}
    \Bigl(n_{1:2n-1,s }-\frac{1}{2}\Bigr)
    \Bigl(n_{2:2n-1,s'}-\frac{1}{2}\Bigr)
   \nonumber \\
   &&\!\!\!\!\!\!\!\qquad
   +\frac{K}{2}\sum_{l=1}^2\sum_{n=1}^N
     \big(r_{l:2n-1}^2+r_{l:2n}^2\big),
   \label{E:H}
\end{eqnarray}
where $c_{l:j,s}^\dagger$ and $c_{l:j,s}$
($c_{l:j,s}^\dagger c_{l:j,s}\equiv n_{l:j,s}$)
create and annihilate, respectively, a $\pi$ electron of spin
$s=\uparrow,\downarrow\equiv\pm$ at site $j$ on chain $l$, while
$r_{l:j}$ denotes the bond distortion caused by the $j$th and $(j+1)$th
carbon atoms on the $l$th chain, on the description of polyacene as a
couple of the {\it trans} isomers of polyacetylene with interchain bonding
at every other site.
The Coulomb interactions, ranging over neighboring sites, are modeled in
$V_{\parallel(\perp)}=U/\kappa\sqrt{1+0.6117a_{\parallel(\perp)}^2}$
(Ref. \onlinecite{O219}), where $\kappa$ is a dielectric parameter, while
$a_{\parallel}$ and $a_{\perp}$ are the average lengths in $\mbox{\AA}$ of
neighboring C$-$C leg and rung bonds, respectively.
Keeping the screened Coulomb parameters in mind, which were initiated by
Chandross and Mazumdar \cite{C1497} and successfully applied to
oligoacenes by Sony and Shukla, \cite{S155208} we adopt $U=8.0\,\mbox{eV}$
with $\kappa=2.0$.
Though the Coulomb interactions in our use are not infinitely ranged but
cut off, such a parametrization must be suggestive and convincing under
long-range correlations of merely moderate effect. \cite{R155204}
The intrachain and interchain electron hoppings are described by
$t_{\parallel}$ and $t_{\perp}$, respectively.
We take $2.4\,\mbox{eV}$ for $t_{\parallel}$ (Ref. \onlinecite{S155208})
and set $t_{\perp}$ equal to $0.864t_{\parallel}$
(Refs. \onlinecite{W341,A2023,A467,Z53}), considering that
$a_{\parallel}\simeq 1.4\,\mbox{\AA}<a_{\perp}\simeq 1.5\,\mbox{\AA}$
(Refs. \onlinecite{B7416,W5720}).
$\alpha$ characterizes the electron-lattice coupling with $K$ being the
$\sigma$-bond elastic constant.
We use, unless otherwise noted, $\alpha=4.1\,\mbox{eV}/\mbox{\AA}$ and
$K=15.5\,\mbox{eV}/\mbox{\AA}^2$
(Refs. \onlinecite{S1269,L77,L477,W341,A2023,A467,Z53}).

   In order to calculate the polarized optical conductivity spectra of
polyacene, we define current operators along the long and short axes as
\begin{eqnarray}
   &&\!\!\!\!\!\!\!\!\!
   {\cal J}_{\parallel}=
    \frac{\sqrt{3}iea_{\parallel}}{2\hbar}\sum_{l,n,s}
    \big[
     (t_{\parallel}-\alpha r_{l:2n-1})
     c_{l:2n-1,s}^\dagger c_{l:2n,s}
   \nonumber \\
   &&\!\!\!\!\!\!\!\!\!\qquad
    +(t_{\parallel}-\alpha r_{l:2n})
     c_{l:2n,s}^\dagger c_{l:2n+1,s}
    -{\rm H.c.}
     \big],
   \\
   &&\!\!\!\!\!\!\!\!\!
   {\cal J}_{\perp}=
    \frac{iea_{\parallel}}{2\hbar}\sum_{l,n,s}(-1)^l
    \big[
     (t_{\parallel}-\alpha r_{l:2n-1})
     c_{l:2n-1,s}^\dagger c_{l:2n,s}
   \nonumber \\
   &&\!\!\!\!\!\!\!\!\!\qquad
    -(t_{\parallel}-\alpha r_{l:2n})
     c_{l:2n,s}^\dagger c_{l:2n+1,s}
    -{\rm H.c.}
     \big]
   \nonumber \\
   &&\!\!\!\!\!\!\!\!\!\qquad
   +\frac{iea_{\perp}}{\hbar}\sum_{n,s}
    t_{\perp}\big(c_{1:2n-1,s}^\dagger c_{2:2n-1,s}-{\rm H.c.}\big).
\end{eqnarray}
Since the charge-transfer excitation energy is of eV order, the system
effectively lies in the ground state at room temperature.
Then the real part of the optical conductivity reads
\begin{equation}
   \sigma_{\parallel,\perp}(\omega)
    =\frac{\pi}{\omega}\sum_i
   |\langle E_i|{\cal J}_{\parallel,\perp}|E_0\rangle|^2
   \delta(E_i-E_0-\hbar\omega),
   \label{E:OC}
\end{equation}
where $|E_i\rangle$ denotes a wave vector of the $i$th-lying state of
energy $E_i$.
The state vectors are calculated within and beyond the HF scheme, being
generally defined as
\begin{eqnarray}
   &&\!\!\!\!\!\!\!
   |E_i\rangle
   =|E_0\rangle^{\quad}_{\rm HF}
   +\sum_{m(k,\mu,s)=1}^{4N}\,\sum_{m(k,\nu,s)=4N+1}^{8N}
   \nonumber \\
   &&\!\!\!\!\!\!\!\quad\times
    f(k,\mu,\nu,s;i)a_{m(k,\nu,s)}^\dagger a_{m(k,\mu,s)}
    |E_0\rangle^{\quad}_{\rm HF}
   \nonumber \\
   &&\!\!\!\!\!\!\!\quad
   +\sum_{m(k_1,\mu_1,s_1)>m(k_2,\mu_2,s_2)=1}^{4N}\,
    \sum_{m(k_1,\nu_1,s_1)>m(k_2,\nu_2,s_2)=4N+1}^{8N}
   \nonumber \\
   &&\!\!\!\!\!\!\!\quad\times
    f(k_1,k_2,\mu_1,\mu_2,\nu_1,\nu_2,s_1,s_2;i)
    a_{m(k_1,\nu_1,s_1)}^\dagger a_{m(k_2,\nu_2,s_2)}^\dagger
   \nonumber \\
   &&\!\!\!\!\!\!\!\quad\times
    a_{m(k_1,\mu_1,s_1)} a_{m(k_2,\mu_2,s_2)}
    |E_0\rangle^{\quad}_{\rm HF},
\end{eqnarray}
where
$|E_0\rangle^{\quad}_{\rm HF}
 \equiv\prod_{m=1}^{4N} a_{m}^\dagger|0\rangle$
is the ground-state HF wave function with $|0\rangle$ being the true
electron vacuum and $a_m^\dagger$ creating an electron in the $m$th HF
orbital of energy $\varepsilon_m$.
The orbital label $m$ is a function of momentum $k$, band label $\lambda$,
and spin $s$, which are all good quantum numbers here.
Any transition of finite momentum transfer is optically forbidden, which
serves to reduce the number of configurations to take into calculation.
Every excited state of the HF type is a single Slater determinant,
\cite{O045122} where $f(k,\mu,\nu,s;i)=\delta_{k\mu\nu s,i}$,
$f(k_1,k_2,\mu_1,\mu_2,\nu_1,\nu_2,s_1,s_2;i)=0$, and thus
$E_i=\!\!\!\!\!\!\quad^{\quad}_{\rm HF\!}\langle E_0|
     {\cal H}|E_0\rangle^{\quad}_{\rm HF}
    -\varepsilon_{m(k,\mu,s)}+\varepsilon_{m(k,\nu,s)}$.
Those of the CI type consist of resonating Slater determinants,
\cite{Y235116} where the coefficients are determined so as to diagonalize
the original Hamiltonian (\ref{E:H}).
Within the single-excitations CI (SCI) scheme,
$f(k_1,k_2,\mu_1,\mu_2,\nu_1,\nu_2,s_1,s_2;i)$ remains vanishing and
therefore, no excited-state Slater determinant mixes with the ground-state
one: $|E_0\rangle=|E_0\rangle^{\quad}_{\rm HF}$.
When we proceed to the single-double-excitations CI (SDCI) scheme, there
occurs a significant correction to the ground-state energy as well as to
every excited-state one.
The CI method enables us to systematically investigate many-body effects
beyond the HF approximation in fairly large systems that we can hardly
diagonalize directly.
It was indeed successfully applied to oligoacenes
\cite{S155208} and related phenyl-based conjugated polymers
\cite{G12763,S125204,S245203,S165218,S177,S165204} with its expansion
truncated at varying level.

\section{Ground-State Phase Competition}

   Even though we take particular interest in the optical features of
polyacene as a Peierls insulator, it is still important for us to have a
bird's-eye view of its competing ground states.
We calculate their energies at the HF and SDCI levels and visualize the
ground-state phase competition in Fig. \ref{F:PhD}.
Besides Peierls-distorted structures, which are here referred to as
in-phase (IP) and out-of-phase (OP) bond-order-wave (BOW) states,
we find two undistorted structures, a metallic state of aromatic (AM)
configuration and a Mott-insulating state of antiferromagnetic (AF)
configuration.
AM is fully symmetric and has no band gap at the HF level.
There opens a gap in the SDCI scheme.
AF is more stabilized and wider gapped with increasing on-site Coulomb
repulsion.
There is another undistorted structure possible, which is characterized
as a charge density wave (CDW), \cite{K305} but it is not stabilized into
the ground state under the screened-Coulomb parametrization.
CDW is realized when we adopt a standard-Coulomb parametrization,
\cite{S155208} which is, however, inferior to reproduce experimental
findings.
\cite{G12763,S125204,S245203,S165218,S177,S165204}

   The HF energies of IP BOW and OP BOW are the same.
IP BOW gains more correlation energy than OP BOW in the SDCI scheme.
However, their energy difference is small and decreases with increasing
conjugation length.
IP BOW and OP BOW are highly degenerate with each other in sufficiently
long acene chains.
They are stabilized conditionally, that is, depending on the
electron-lattice coupling strength, against AM and AF under weak and
strong electron-electron correlations, respectively.
Moderate Coulomb interactions enhance the Peierls instability.
The realistic parameters in our use, which are indicated with $\times$ in
Fig. \ref{F:PhD}, sit in close vicinity to a phase boundary.
The ground state is a Mott insulator, but it closely competes with
Peierls insulators.
Many-body electron correlations seem to contribute toward a closer
competition between them.
They are very much likely to coexist in polyacene. \cite{C473}
Then, how can we distinguish between the two distorted structures?
\begin{figure*}
\centering
\includegraphics[width=168mm]{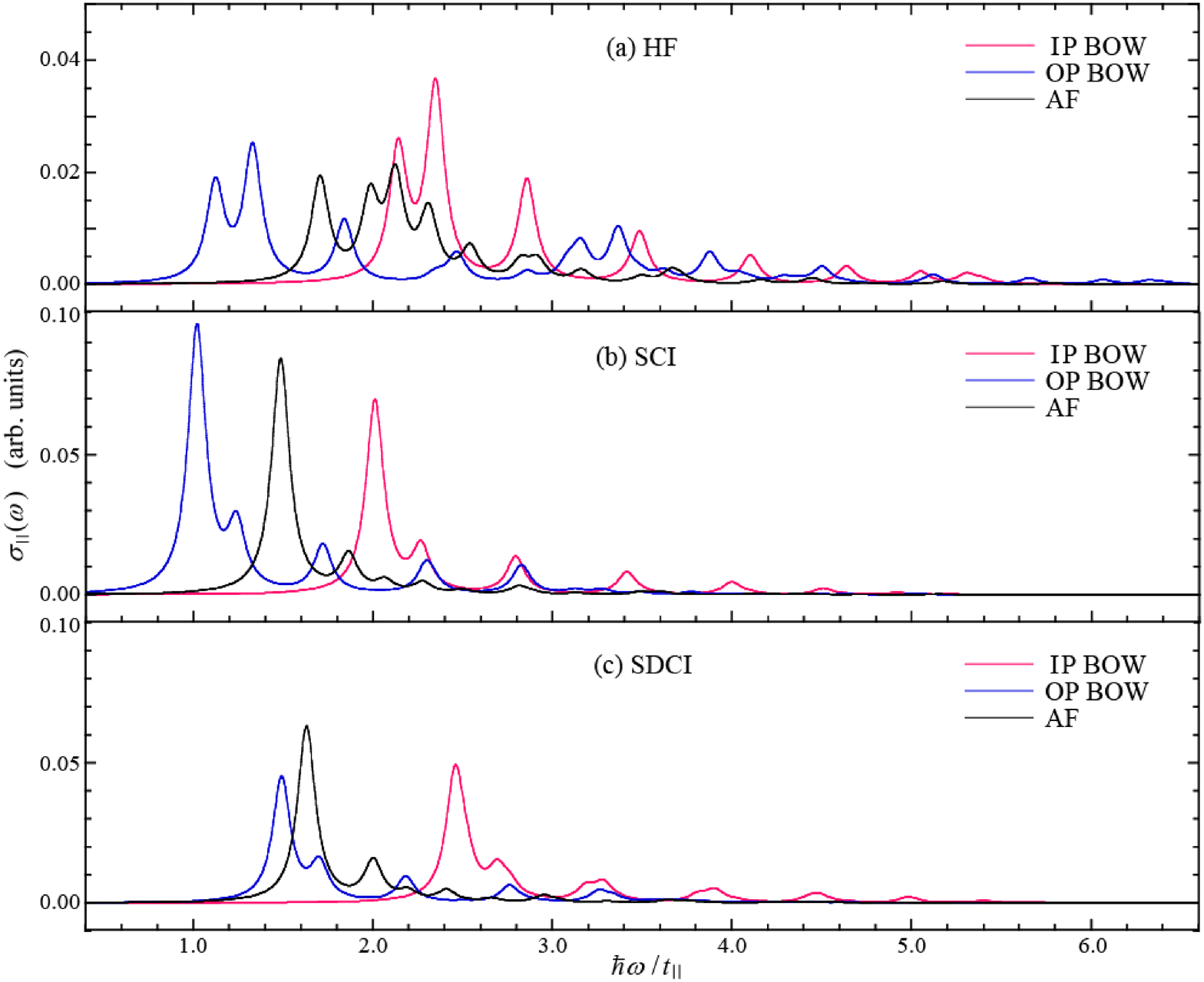}
\vspace*{-2.5mm}
\caption{(Color online)
         HF, SCI, and SDCI calculations of the long-axis-polarized
         optical conductivity spectrum for IP BOW, OP BOW, and AF at
         $N=16$ under the screened-Coulomb and moderate-coupling
         parameters.
         Every spectral line is Lorentzian broadened by a width of
         $0.06t_{\parallel}$.}
\label{F:ScOC}
\end{figure*}

\begin{figure*}
\centering
\includegraphics[width=168mm]{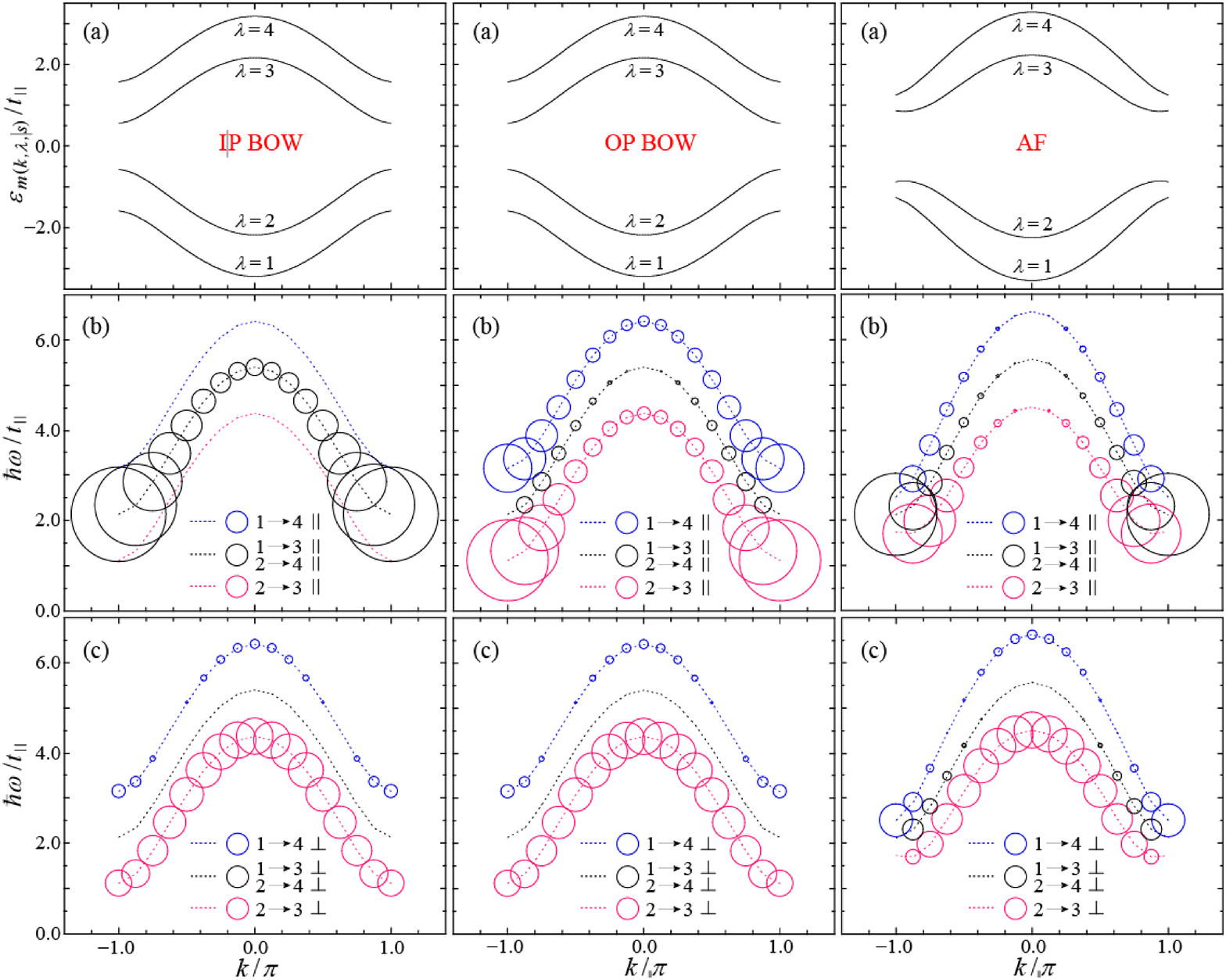}
\vspace*{-2.5mm}
\caption{(Color online)
         HF calculations of the dispersion relations of the $\pi$-electron
         valence ($\lambda=1,2$) and conduction ($\lambda=3,4$) bands (a)
         and the ``momentum-resolved" polarized optical conductivity
         parallel (b) and perpendicular (c) to the long axis for IP BOW,
         OP BOW, and AF at $N=16$ under the screened-Coulomb and
         moderate-coupling parameters, where the area of each circle
         corresponds to the spectral weight.}
\label{F:ScOCDTL}
\end{figure*}

\section{Optical conductivity spectra}

   We calculate the polarized optical conductivity spectra of
energetically competing IP BOW, OP BOW, and AF with varying number of
contributive Slater determinants and compare them in Fig. \ref{F:ScOC}.
First of all we are impressed with an optical contrast between IP BOW and
OP BOW, which is noticeable even in the HF scheme.
They are degenerate with each other in their energetics but
distinguishable from each other in their optics.
Their most intense peaks appear far apart and in between is that of AF.
The HF description of the relative intensity and position of each peak is
much poorer than the CI findings, but it is useful enough to illuminate
the optical nature of individual phases in a qualitative manner.

   Based on the HF energy scheme, Fig. \ref{F:ScOCDTL} analyzes the
polarized optical conductivity spectra parallel and perpendicular to the
conjugation direction.
There are four molecular orbitals in each unit cell and they are molded
into two valence and two conduction bands fulfilling the electron-hole
symmetry.
The dispersion relations of IP BOW and OP BOW are exactly the same within
the HF scheme and remain alike even with many-body Coulomb correlations
fully considered. \cite{R155204}
Their dipole transition matrices are also the same provided the excitation
light is polarized in the rung direction, as is shown in
Fig. \ref{F:ScOCDTL}(c), where most of the spectral weight comes from
$2$-to-$3$ interband transitions.
It is due to the parity-definite molecular orbitals \cite{K97,K7236}
that both $1$-to-$3$ and $2$-to-$4$ interband excitations make no
contribution to the rung-direction optical conductivity.
Under the reflection about the plane bisecting every rung bond, the
valence and conduction bands labeled $1$ and $3$ are of symmetric
character, whereas those labeled $2$ and $4$ are of antisymmetric
character, in both IP BOW and OP BOW.
Any transition with parity unchanged is optically forbidden.
On the other hand, photoirradiation in the conjugation direction reveals
a striking contrast between IP BOW and OP BOW, as is shown in
Fig. \ref{F:ScOCDTL}(b), where all the spectral weight comes from
$1$-to-$3$ and $2$-to-$4$ interband transitions in IP BOW, while little
contribution from them in OP BOW.
The conjugation-direction optical conductivity on an OP-BOW background
arises mostly from $2$-to-$3$ and $1$-to-$4$ excitations.

   The $\sigma_{\parallel}(\omega)$ spectra of IP BOW and OP BOW are thus
distinguishable.
Their main peaks sandwich the AF low-energy absorption bands
(Fig. \ref{F:ScOC}).
With Coulomb correlations fully taken into account, the most intense
absorption band is sharpened and grown up in general.
Those of OP BOW and AF are close to each other, staying away from that
of IP BOW.
The screened-Coulomb and moderate-coupling parameters in our use stabilize
AF slightly more than IP BOW.
Band gaps are barometers of stabilization and therefore AF is gapped wider
than IP BOW [Fig. \ref{F:ScOCDTL}(a)].
Although the BOW Peierls gap is smaller than the AF Mott-Hubbard gap, the
main absorption peak in the AF spectrum appears much below that in the
IP-BOW one.
This trick is due to the optically forbidden lowest-lying Peierls-gap
excitation through a long-axis-polarized photon on an IP-BOW background.
\begin{figure*}
\centering
\includegraphics[width=116mm]{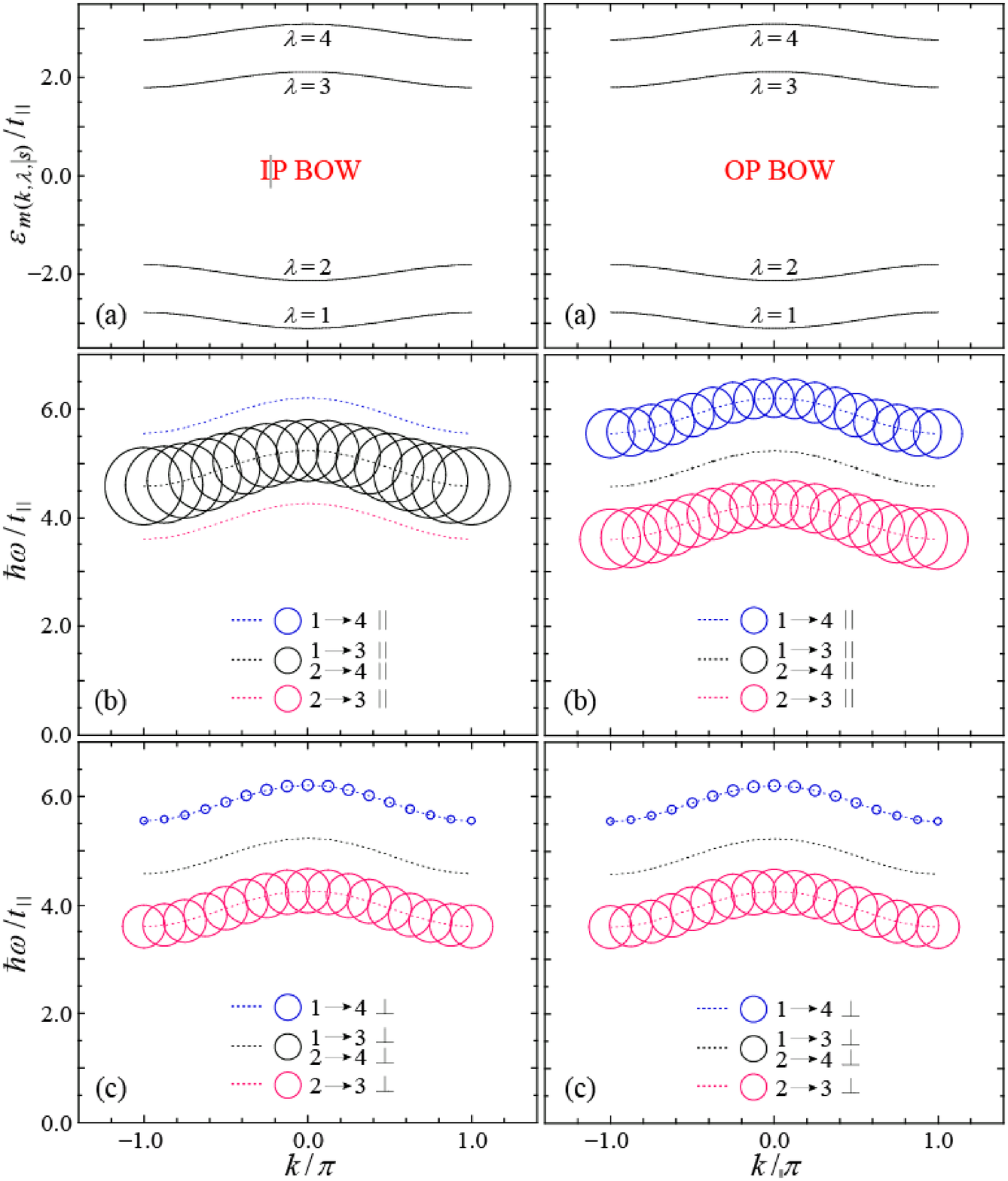}
\vspace*{-2.5mm}
\caption{(Color online)
         The same as Fig. \ref{F:ScOCDTL} but the screened-Coulomb and
         strong-coupling parameters.}
\label{F:SKScOCDTL}
\end{figure*}

\begin{figure*}
\centering
\includegraphics[width=148mm]{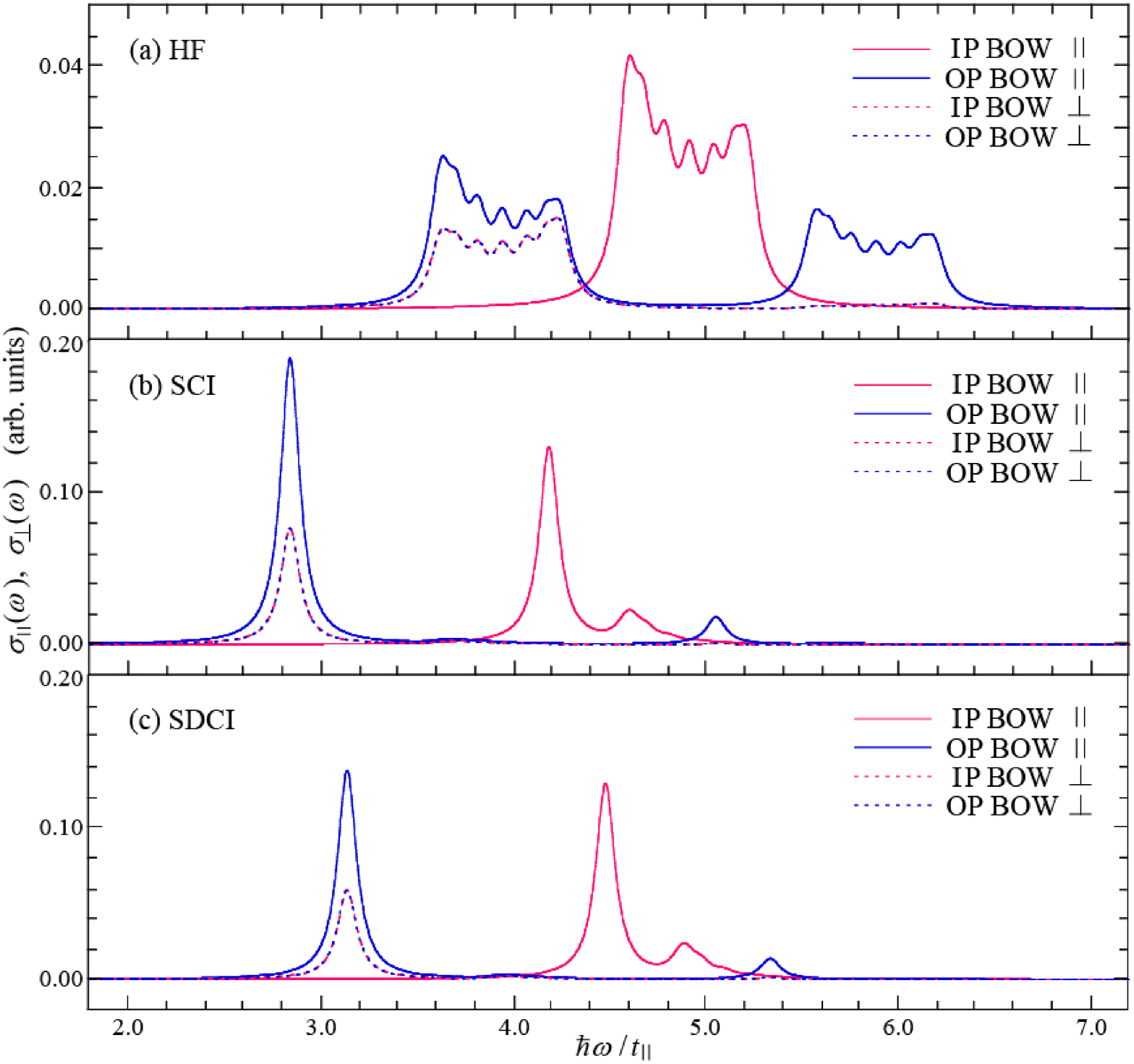}
\vspace*{-2.5mm}
\caption{(Color online)
         HF, SCI, and SDCI calculations of the polarized optical
         conductivity spectra parallel ($\parallel$) and perpendicular
         ($\perp$) to the long axis for IP BOW and OP BOW at $N=16$ under
         the screened-Coulomb and strong-coupling parameters.
         Every spectral line is Lorentzian broadened by a width of
         $0.06t_{\parallel}$.}
\label{F:SKScOC}
\end{figure*}

\begin{figure*}
\centering
\includegraphics[width=148mm]{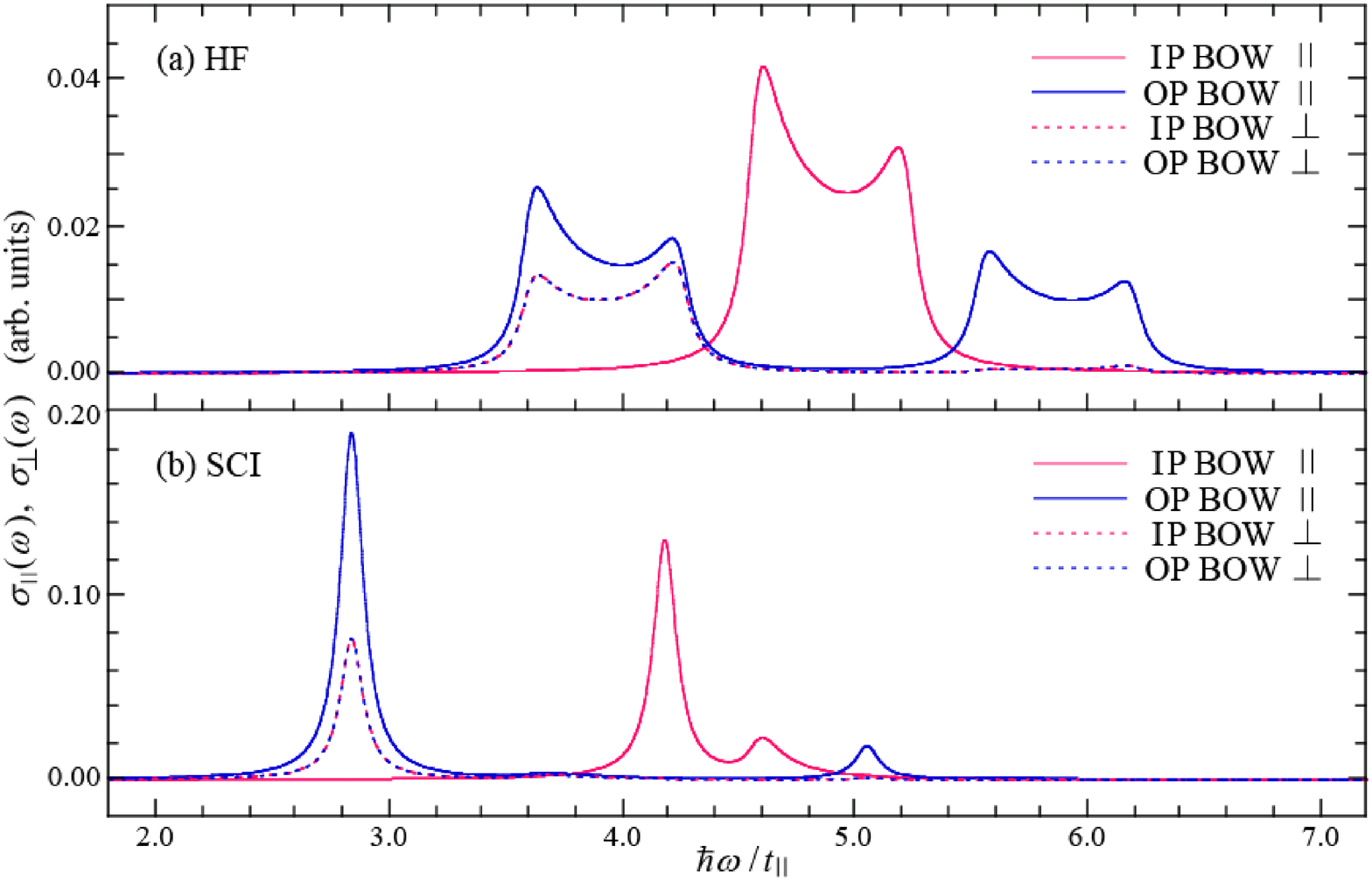}
\vspace*{-2.5mm}
\caption{(Color online)
         The same as Fig. \ref{F:SKScOC} but $N=128$, where the SDCI
         scheme is never feasible.}
\label{F:SKScOCLRG}
\end{figure*}

   The optical contrast between IP BOW and OP BOW is more and more
accentuated with increasing coupling strength.
If we reduce the elastic constant $K$ to half the present value, as is
indicated with $\circ$ in Fig. \ref{F:PhD}, the electron-lattice coupling
is effectively strengthened and any distorted structure is sufficiently
stabilized against the Mott insulator.
It may be the case, for instance, with organic molecular compounds of
alternating tetrathiafulvalene and chloranil \cite{I060302} and/or
halogen-bridged transition-metal linear-chain complexes.
\cite{A1415,I241102}
Figure \ref{F:SKScOCDTL} shows the same calculations as in
Fig. \ref{F:ScOCDTL} but $K=7.75\,\mbox{eV}/\mbox{\AA}^2$.
In the case of electron-lattice interactions predominating over
electron-electron correlations, both valence and conduction bands are much
less dispersive in general and the long-axis-polarized dipole transitions
between them are really eloquent of their background lattice distortion.
In the conjugation direction, $1$-to-$4$ and $2$-to-$3$ dipole matrix
elements vanish on an IP-BOW background, while those of $1$-to-$3$ and
$2$-to-$4$ are negligibly small on an OP-BOW background.
For light polarized in the rung direction, IP BOW and OP BOW similarly
behave and their optical features are essentially given by $2$-to-$3$
interband transitions.

   We present in Fig. \ref{F:SKScOC} the resultant polarized optical
conductivity spectra.
For light polarized in the conjugation direction, IP BOW and OP BOW
exhibit a single and two well-separate absorption bands, respectively.
With configuration interactions fully taken into account, the single
IP-BOW absorption band consists of an intense peak and its satellite in
the high-energy side, whereas a couple of the OP-BOW absorption bands are
both single-peaked, in a practical sense, due to the predominant
zone-center excitations.
Many-body electron correlations further have a significant effect on the
relative intensity of the absorption bands.
The spectral weight of the higher-energy OP-BOW absorption band turns out
much smaller than that of the lower-energy one.
For light polarized in the rung direction, IP BOW and OP BOW are
degenerate with each other.
Their spectra slightly differ beyond the HF scheme, but the difference is
hardly recognizable in practice.
The common $\sigma_{\perp}(\omega)$ spectrum is peaked similarly to the
IP-BOW $\sigma_{\parallel}(\omega)$ spectrum, but high-energy absorptions
are strongly suppressed in the rung direction.

   Finally we stress that the present findings are not individual features
of small oligoacenes but symbolize polyacene in the thermodynamic limit.
All the features but irregular outlines peculiar to small clusters in
Fig. \ref{F:SKScOC} remain unchanged with further increasing conjugation
length, as is demonstrated in Fig. \ref{F:SKScOCLRG}.
Indeed the absolute location and the relative intensity of every
absorption depend on the number of benzene rings, \cite{S155208}
but they are almost converging at $N=16$.

\section{Summary}

   We have optically characterized competing ground states of polyacene
and revealed a striking contrast between the two highly-degenerate
Peierls-distorted states in particular.
The {\it cis-} and {\it trans-}distorted structures, which we refer to as
IP BOW and OP BOW, are almost degenerate in their energetics but quite
distinct in their optics.
The highest-occupied-molecular-orbital
(HOMO)-to-lowest-unoccupied-molecular-orbital (LUMO) transition through
a long-axis-polarized photon is allowed in OP BOW but forbidden in IP BOW.
There appear two well-separate absorption bands against a background of
OP BOW and in between does a single absorption band with an IP-BOW
background sit.
Such distinct features may not be demonstrated as they are in polyacene
of predominantly strong electron-electron correlation, but with growing
electron-lattice coupling they become detectable literally.
We may consider heteroacenes such as paracyanogen, \cite{W23} 
as well as substituting hydrogen atoms in polyacene with larger molecules,
in an attempt to tune the elastic properties and to realize a
coupling-dominant situation.

   The realistic set of parameters is located in close vicinity to a phase
boundary, where an antiferromagnetic Mott-insulating state, which we refer
to as AF, is slightly lower in energy and gapped wider than the two
Peierls-distorted states.
The most intense AF absorption peak nevertheless appears far below the
IP-BOW absorption band for light polarized in the conjugation direction.
For light polarized in the rung direction, on the other hand, the common
optical gap of IP BOW and OP BOW is naively smaller than that of AF.
The relative location of their main absorption peaks remains qualitatively
unchanged with varying configuration interactions, though the excitonic
effects on the optical spectra are significant in general.

   There is another example \cite{Y235116} of optically characterizing
distinct ground states in competition.
Some class of platinum-halide ladder compounds exhibits Peierls-distorted
CDW ground states of the IP and OP types.
According to their interchain valence arrangements, their optical
conductivity spectra are differently peaked.
In this case, however, IP CDW and OP CDW not only look different in their
energetics but also possibly present some contrast for the Raman
spectroscopy.
In the present case, IP BOW and OP BOW are so degenerate with each other
that much effort has been devoted to solving the problem of which is
energetically preferable. \cite{B4891}
Then the optical contrast between them will indeed come in useful for
identifying them.
Regular substitution of carbon atoms in polyacene, for instance, with
nitrogen atoms, may lead to further stabilization of the Peierls-distorted
structures against the aromatic one.
IP BOW and OP BOW remain closely competing with each other in
polypyridinopyridine. \cite{W23}
We hope our calculations will stimulate a renewed interest in polycyclic
aromatic hydrocarbons.

\acknowledgments

   The author is grateful to J. Ohara for fruitful discussion.
This work was supported by the Ministry of Education, Culture, Sports,
Science, and Technology of Japan.


\end{document}